\RequirePackage{fix-cm}
\documentclass{svjour3}                     
\smartqed  

\usepackage{amsmath}    
\usepackage{amsfonts}  
\usepackage{graphicx}   
\usepackage{bm}  

\journalname{Journal of Low Temperature Physics}

\begin{document}
\title{Cooling by Thermodynamic Induction} 

\author{S. N. Patitsas}
\institute{University of Lethbridge,\\4401 University Drive, Lethbridge AB, Canada, T1K3M4}

\date{Received: date / Accepted: date}

\maketitle
	
\begin{abstract}
A method is described for cooling conductive channels to below ambient temperature.  The thermodynamic induction principle dictates that the electrically biased channel will cool if the electrical conductance decreases with temperature.  The extent of this cooling is calculated in detail for both case of ballistic and conventional transport with specific calculations for carbon nanotubes and conventional metals, followed by discussions for semiconductors, graphene, and metal-insulator transition systems.  A theorem is established for ballistic transport stating that net cooling is not possible.   For conventional transport net cooling is possible over a broad temperature range, with the range being size-dependent.  A temperature clamping scheme for establishing a metastable nonequilibrium stationary state is detailed and followed with discussion of possible applications to on-chip thermoelectric cooling in integrated circuitry and quantum computer systems.

\end{abstract}

%

\maketitle 

\section{Introduction} \label{sec:intro}

Many major advances in physics have come about after the development of new cooling techniques.  Examples include the discovery of superconductivity and superfluidity after the development of techniques for liquifying gases such as hydrogen, oxygen, and helium, and more recently the discovery of gaseous Bose-Einstein condensation after the techniques of Doppler cooling and evaporative cooling were developed~\cite{Pobell,Schawlow1975,Wineland1978,Wieman1995,Ketterle1996}.  

Here, I present a new way to achieve cooling which exploits the thermodynamic induction (TI) effect.  How this induction arises in a dissipative nonequilibrium system has been described previously~\cite{Patitsas2014,Patitsas2015}.  In such systems, certain subsystems which have a gating property can be induced away from thermodynamic equilibrium in unexpected ways~\cite{Patitsas2014}.  The rate of entropy generation plays an important role in the theory and the analysis shows that there exists an important energy scale given by the product of the rate of entropy production, the characteristic time scale for fluctuations, and the temperature.  This energy scale influences the system dynamics and gives rise to TI.  In fact for certain far-from-equilibrium situations this energy scale exceeds all of the other energy scales, the direct Hamiltonian approach for describing the system dynamics is difficult to implement, and the dissipative dynamical approach based on entropy production provides the best approach for describing the induction effects that emerge in these systems.  For example, entropy production principles are believed to play an important role in the way individual atoms are manipulated by the tip of a STM~\cite{Patitsas2015}.    


I show that in certain thermoelectric systems the gate can be induced to achieve lower temperature than otherwise expected.  This effect is not described within linear nonequilibrium theory and may be thought of as a nonlinear complement to traditional Peltier cooling~\cite{AM}.  It is worth exploring if TI can be exploited towards achieving a practical method for cooling.  Results presented here indicate that this may indeed be the case for certain systems.  In Section~\ref{sec:TItheory} the basic theory for TI is briefly reviewed, before carefully describing the physical system in Section~\ref{sec:TS}.  This leads to the main stationary state result presented in Section~\ref{sec:statst}.  The case of pure ballistic transport is treated in Section~\ref{sec:ball} with numerical calculations on zigzag carbon nanotubes, and then conventional transport with calculations for highly purified silver and copper samples are presented in Section~\ref{sec:conv}.  Finally I discuss my findings and conclude.

\section{Thermodynamic Induction Theory}  \label{sec:TItheory}

  For general TI theory, one may refer to \cite{Patitsas2014} for the case of $n$ variables and for the continuum limit see Ref.~\cite{Patitsas2015}.  For the theoretical development here, a discrete 2-variable approach will suffice for the most part, so this will be briefly developed here.
I emphasize that the dynamical equations describing the approach to equilibrium for these two variables are uncoupled to first order, i.e., the two Onsager coefficients $L_{12}=L_{21}$ are zero.  It is in the second order with nonlinear terms that TI effects spring forth and create the interesting coupling that is at the heart of this work.

These variables $x_1$ and $x_2$ are thermodynamic in nature with equilibrium values $x_{1_0}$ and $x_{2_0}$.
  The difference variables $a_1=x_1-x_{1_0}$ and $a_2=x_2-x_{2_0}$ are used to make an expansion of the total entropy $S_T$ with small deviations from equilibrium as
\begin{equation}
\Delta S_T\equiv S_T(a_1,a_2)-S_T(0,0)= -\frac{1}{2}\sum_{p=1}^2\sum_{q=1}^2 c_{pq}^{-1}a_p a_q ~.    \label{quadST}
\end{equation}	
	 Generalized thermodynamic forces (affinities) are defined by $X_i=\frac{\partial \Delta S_T}{\partial a_i}$~\cite{mazur}.  These conjugate parameters are interrelated as:    
\begin{equation}
a_i=-\sum_{j=1}^2 c_{ij}X_j       ~  ,    \label{agX}
\end{equation} 
where $c_{12}=c_{21}$, i.e., the generalized capacitance matrix $c_{ij}$ is symmetric and positive definite, thus ensuring thermodynamic stability.  These matrix elements also govern the extent of fluctuations~\cite{mazur}.  For example, in the decoupled case where $c_{12}=0$ then $k_B c_{22}$ is the root-mean-square deviation of the probability distribution for $a_2$, i.e, 
\begin{equation}
P_0(a_2)=c~\exp{\left(\frac{-a_2^2}{2k_B c_{22}}\right)}~\, ,   \label{P0}
\end{equation}
where $c$ is a normalization constant.
If one or both of these variables is pushed away from equilibrium then the dynamics is concerned with the approach back to equilibrium.  Such dynamics are often governed by transport processes such as thermal conductivity, electrical conductivity, etc.  These transport processes are irreversible:  the total system entropy increases with time as the system approaches equilibrium.  Note that
\begin{equation}
\frac{dS_T}{dt}\equiv\dot{S_T}= \sum_{p=1}^2 X_p \dot{a}_p    ~.  \label{STdot}
\end{equation}
		
	In Eq.~(\ref{STdot}) the time derivative $\dot{a}_p$ must be handled with care.	A proper dynamical approach uses coarse-grained averages of the time derivatives.  For variable $x_i$ the time step $\Delta t_i$ chosen for the dynamics must be much larger than the relevant correlation time $\tau_i^*$ for fluctuations.  Fluctuations play an important role and are the ultimate source of TI.  They are also the fastest process in this analysis, with $\tau_i^*$ being a measure of the width of the fluctuation force correlation function, typically very short~\cite{Reif}.  
We closely follow the notation in Ref.~\cite{mazur}, and use a bar to denote the coarse-grained time derivative as $\bar{\dot{a}}_i$.  The definition is
\begin{equation}  
\bar{\dot{a}}_i\equiv\frac{1}{\Delta t_i}\int_t^{t+\Delta t_i}{\langle\dot{a}_i\rangle dt'}=\frac{1}{\Delta t_i}\langle a_i(t+\Delta t_i)-a_i(t)\rangle  ~  .
\end{equation}
Note that the symbols $\langle~\rangle$ denote ensemble averages over accessible microstates of the system.  Addition of the subscript, 0, refers to equilibrium states in particular.

To evaluate this coarse-grained derivative I closely follow the approach in Ref.~\cite{Reif}.  In this approach, a nonequilibrium average value is determined by evaluating an equilibrium average accompanied by the insertion of the ubiquitous $\exp(\Delta S_T/k_B)$ probability weighting factor~\cite{mazur,LandauL}.  We emphasize that this must be the total system entropy change.  Pedagogically, this approach is known to establish the Onsager symmetry relations~\cite{Reif}. 
Explicitly then:  
\begin{equation}
\langle\dot{a}_i(t')\rangle=\left\langle\dot{a}_i(t')e^{\frac{\Delta S_T(t'-t)}{k_B}}\right\rangle_0~,   \label{expS}          
\end{equation}
where $\Delta S_T(t'-t)\equiv S_T(t')-S_T(t)$. 
The physical interpretation is simple;  For a given state at time $t$, every possible outcome at time $t'$ is considered with each outcome statistically weighted by the number of accessible system microstates. 
 Since the change in entropy is small, a linear expansion is warranted, leaving:
\begin{equation}
\langle\dot{a}_i(t')\rangle=\frac{1}{k_B}{\langle\dot{a}_i(t')\Delta S_T(t'-t)}\rangle_0~,   \label{adot2b}
\end{equation}
since $\langle \dot{a}_i \rangle_0=0$.  Integrating both sides of Eq.~(\ref{adot2b}) over the time interval $\Delta t_i$ gives the coarse-grained time derivative:
\begin{equation}
\bar{\dot{a}}_{i}=\frac{1}{k_B\Delta t_i}\int_{t}^{t+\Delta t_i}{dt'\int_t^{t'}{dt''\langle\dot{a}_i(t')\dot{S_T}(t'')\rangle_0}}\, . \label{adot4}
\end{equation}

At this point a distinction is made between the two variables.  Variable 1 will be designated as the dynamical reservoir (DR) and variable 2 is the gate (GT) variable.  The DR variable $x_{DR}$ approaches equilibrium slowly and with a virtually limitless supply.  An example is the charge variable in an RC circuit with a very large capacitance.  
  The gate variable $x_{GT}$ controls the slower DR variable by having the conductance of the DR depend on the state of the gate.
  Following Ref.~\cite{Patitsas2014} the dynamics for the DR is straightforward and given by
\begin{equation}
\dot{a}_{DR}=M_{11}X_{DR} \, . \label{adotDR1}
\end{equation}
By Eqs.~(\ref{STdot}) and (\ref{adotDR1}), entropy is created by the DR at a rate $\dot{S}_{DR}=M_{11}X_{DR}^2$.  If $M_{11}$ is constant then this is a standard result~\cite{Reif}.  Also, as mentioned above, a linear coupling term of the form $L_{12}X_{GT}$ is not present.  If it were present, then the nonlinear analysis presented below would merely provide small, second order corrections to the dynamics.  It cannot be emphasized enough that the analysis presented below will provide a potent nonlinear dynamical coupling between otherwise uncoupled variables.  Here, it is not the case that $M_{11}$ is constant, and $M_{11}$ depends on the gate variable $x_2=x_{GT}$.  The simplest way to express this dependence is
\begin{equation}
M_{11}=L_{11}+\gamma a_{GT} \, , \label{M11}
\end{equation}
where $L_{11}$ and $\gamma$ are constants.  Using Eq.~(\ref{STdot}) the total rate of entropy production has contribution given by $L_{11}X_{DR}^2+L_{22}X_{GT}^2$~\cite{mazur}, as well a term coming from the nonlinear part of Eq.~(\ref{M11}) which is given by:
\begin{equation}
\dot{S}_{T_{nonlin}}= \gamma X_{DR}^2 a_{GT}  \, .  \label{STdotgamma}
\end{equation}
When applying Eq.~(\ref{adot4}) to the gate variable, one recovers the expected linear term $L_{22}X_2$, but by implementing Eq.~(\ref{STdotgamma}) one also finds~\cite{Patitsas2014} the interesting induction term, to give
\begin{equation}
{{\dot{a}}}_{GT} = L_{22}X_{GT} + \tau_{GT}^* \gamma L_{22} X_{DR}^2\, . \label{2vardyneq2}
\end{equation}
The fluctuation time $\tau_{GT}^*$ plays an important role throughout this work.  This time comes out of the time integration for the linear term and gets absorbed into the transport coefficient $L_{22}$.  For the nonlinear term the extra time integral gives the extra factor of $\tau_{GT}^*$, prominent in the second term on the right hand side of Eq.~(\ref{2vardyneq2}), and for this reason $\tau_{GT}^*$ finds itself in so many of the equations which follow.

A stationary state for the gate is defined by setting ${{\dot{a}}}_{GT}=0$ which gives
\begin{equation}
X_{GT_{ss}}=- \tau_{GT}^* \gamma X_{DR}^2 \, , \label{XGTSS}
\end{equation}
and 
\begin{equation}
a_{GT_{ss}}= \tau_{GT}^* c_{22}\gamma X_{DR}^2 \, . \label{aGTSS}
\end{equation}
Stationary states have interesting properties.
Compared to the situation with $a_{GT}=0$, entropy production by the DR is increased from the rate $L_{11}X_{DR}^2$ by a positive definite amount 
\begin{equation}
\dot{S}_{extra}= \tau_{GT}^* c_{22}\gamma^2 X_{DR}^4~\, ,   \label{SdotExtra}
\end{equation}
so that
\begin{equation}
\dot{S}_{DR}=M_{11}X_{DR}^2=L_{11}X_{DR}^2+\gamma a_{GT}X_{DR}^2 =L_{11}X_{DR}^2+\dot{S}_{extra}~\, .   \label{SdotDR}
\end{equation}
Also, the entropy of the gate changes from the equilibrium value by an amount:
\begin{equation}
\Delta {S}_{GT}= -\tau_{GT}^* \dot{S}_{extra}~\, .   \label{DeltaS}
\end{equation}
This drop in entropy can mean many things;  Here, under the right conditions, the gate will cool, in the stationary state, to below-ambient temperature.

\section{Thermoelectric System} \label{sec:TS}



\begin{figure}[ht]
	\centering
		\includegraphics[width=0.9\textwidth]{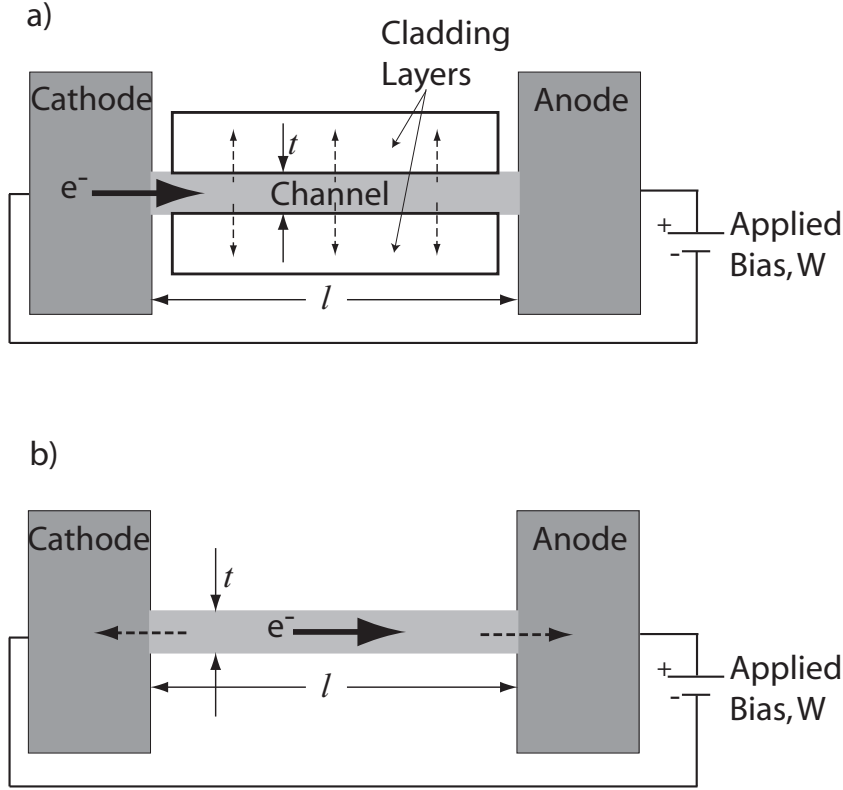}
\caption{\label{fig:plank} Schematic circuit diagrams depicting a conductive channel of length $l$, thickness $t$, connecting a cathode to an anode.  The channel has rectangular cross section $w\times t$ ($w$ into page).  Electron flow associated with the dynamical reservoir is depicted by a thick solid arrow.  Dashed arrows depict the thermodynamically-induced heat flow away from the channel for the case of negative $\gamma$.  In case (a) the heat flow is perpendicular to the electron flow, as heat is deposited into the electrically insulating cladding layers.  In case (b) heat flows from the channel and into the cathode and anode.}
	\label{fig:plank}
\end{figure}


Specifically, I analyze the systems depicted in Fig.~\ref{fig:plank}, each of which experiences two nonequilibrium processes, one concerned with electrical transport and the other with thermal transport.  Process 1, electrical transport through a conductive channel, will play the role of the DR.  More specifically, the DR is not just the physical system of cathode, anode, emf source, and conductive channel, but is also an associated process, here of electronic transport.

	
	The thermal transfer process will have a gating effect on the electrical conductance of the reservoir, simply by having the conductive channel's electrical conductivity, $\sigma$, depend on temperature.  The channel is the most important component of these systems and must also conduct heat into or out of an adjacent heat bath.  The channel connecting anode and cathode across a distance $l$ has rectangular cross section $w\times t$ and is much narrower than the leads, thus forming an electrical bottleneck.  The thickness dimension $t$ is considered to be small compared to the size of the leads, but the depth $w$ is not necessarily small.   Similar to the DR, the gate is not just the physical system of the channel shown in Fig.~\ref{fig:plank}. The gate subsystem is comprised of the channel, the anode and cathode, an associated process of heat transfer, and a thermal bath to absorb or release heat.  In the case of Fig.~\ref{fig:plank}(a) the two electrically-insulating cladding layers compose the bath, while in the case of Fig.~\ref{fig:plank}(b) the channel is free-standing and the cathode and anode together form the bath.   
Qualitatively, the effect of TI is that when bias is applied, heat will flow either into or out of the channel, always in a way as to increase the electrical conductance of the channel.  As pointed out in Refs.~\cite{Patitsas2014,Patitsas2015} the effects of TI are imperceptible in many real systems.  Thus, the analysis presented here focuses on finding conditions where the effects of TI are detectable. 

In this discrete thermodynamic model, $x_{DR}$ is the number of electrons in the anode~\cite{Patitsas2014}.  The thermodynamic force conjugate to this variable is known to be simply related to the change in electron chemical potential $\mu$ relative to the cathode: 
\begin{equation}
X_{DR}\equiv \frac{\partial\Delta S_T}{\partial x_{DR}}=-\frac{\Delta\mu_{DR}}{T}=+\frac{e}{T}W \, ,  \label{Y1toW}
\end{equation}
where $W$ is the applied electric potential difference~\cite{Patitsas2015,mazur}.
Since the electrical current is $I=e\dot{x}_{DR}$ then Eq.~(\ref{adotDR1}) represents Ohm's law and the channel conductance $G$ is related to the Onsager coefficient $M_{11}$ as~\cite{Patitsas2015}:
\begin{equation}
M_{11}=\frac{T}{e^2}G \, .   \label{MtoG}
\end{equation}
Ignored at this level of analysis is the corresponding flow of thermal energy caused by the electric field inside the channel~\cite{AM,mazur}.   Though this is a valid concern, as long as appreciable thermal gradients in the channel are not created, this energy flow does not change the internal energy of the channel in steady state and one can safely ignore it.  Also, it's worth noting that these thermoelectric effects are less significant at lower temperatures~\cite{AM}.
The gating property that is crucial to this analysis involves thermal transport, though in a different way.  This type of thermal transport is indicated by the dashed arrows in Fig.~\ref{fig:plank}.  Heat flow perpendicular to the electrical current is possible if electrically insulating (cladding) material is present directly adjacent to the channel, as shown in Fig.~\ref{fig:plank}(a).  In Fig.~\ref{fig:plank}(b) the direction of induced thermal transport is into or out of the anode and cathode.   Specifically, the variable $x_{GT}$ is the internal energy $U$ of the conducting channel, and $a_{GT}$ is the change from the equilibrium value.  The conjugate force can be related to the temperature difference, $\Delta T$, between the channel and the ambient temperature, $T$, of the electrical leads, as~\cite{Patitsas2015,mazur} 
\begin{equation}
X_{GT} = -\frac{1}{T^2}\Delta T \, .  \label{XG}
\end{equation}
The $c_{22}$ coefficient is directly related to the heat capacity $C_{GT}$ of the channel by the relation: 
\begin{equation}
a_{GT}=-T^2 C_{GT} X_{GT}~.   \label{g22}
\end{equation} 
When this thermal subsystem is out of equilibrium, the thermal flux is given by $L_{22}X_{GT}=-h\Delta T$ where $L_{22}$ is the Onsager coefficient and $h$ is the thermal conductance of the channel~\cite{Reif}.
The relaxation time scale for this thermal transport is
\begin{equation}
\tau_{GT}=\frac{C_{GT} T^2}{L_{22}} = \frac{C_{GT}}{h}~\, ,   \label{tauGT}
\end{equation}
and is not to be confused with the (shorter) fluctuation time scale $\tau_{GT}^*$.
This conventional heat flow, not illustrated in Fig.~\ref{fig:plank}, tends to oppose the induction effect and creates a balance after the stationary state is achieved, as discussed further, below.

The gate and the DR become coupled when $M_{11}$ depends on the gate variable, as described by Eq.~(\ref{M11}). 
Using Eq.~(\ref{MtoG}), the induction coefficient $\gamma$ defined by Eq.~(\ref{M11}) can be expressed as
\begin{equation}
\gamma =\frac{T}{e^2 C_{GT}}\frac{\partial G}{\partial T} ~, \label{gamm112}
\end{equation}
where $c_V$ is the volume specific heat capacity of the channel material.
It is helpful to define a dimensionless version of $\gamma$ as:
\begin{equation}
\kappa\equiv \frac{\partial\ln G}{\partial \ln T} =\frac{e^2 C_{GT}}{G}\gamma . \label{kappa}
\end{equation}

The dynamical equations for the approach to equilibrium are also affected by simple Joule heating of the channel.  For the DR the dynamics are still determined by Eq.~(\ref{adotDR1}).
For the gate, Eq.~(\ref{2vardyneq2}) is modified to become
\begin{equation}
{{\dot{a}}}_{GT} = P_J +L_{22}X_{GT} + \tau_{GT}^* \gamma L_{22} X_{DR}^2\, , \label{2vardyneq3}
\end{equation}
where $P_J=GW^2$ is the Joule heating.  The characteristic time scale $\tau_{GT}^*$ for thermal fluctuations is generally very short~\cite{Patitsas2014}.  
Here we consider only the case where the applied bias $W$ is slowly varying, so the first and third terms in the right-hand side of Eq.~(\ref{2vardyneq3}) may be treated as essentially constant.
The induction term is the third term in the right-hand side of Eq.~(\ref{2vardyneq3}), i.e., the existence of an induced heat flux given by 
\begin{equation}
J_{ind} = e^2 h\tau_{GT}^*\gamma W^2 =  \frac{\tau_{GT}^*}{\tau_{GT}} \kappa P_J \,. \label{Jind}       
\end{equation}
  It is important to note that for negative $\gamma$ the channel is cooled with a power level given by $P_{cool}=-J_{ind}$.  For many systems the ratio, $r$, of gate fluctuation time to relaxation time is very small.  When looking for systems that might exhibit measurable TI cooling one would therefore look at systems with very fast relaxation times.  These systems might typically be small in physical size; In fact for very small systems with a gating characteristic relaxation rates approaching $1/\tau_{GT}^*$ are certainly possible.  In these systems, net cooling, where $P_{cool}$ exceeds $P_J$, is a possibility if $\kappa<-1$.  This important point will be re-visited below in the context of both conventional and ballistic transport.
	

\section{Stationary State}\label{sec:statst}
	
In the stationary state, ${{\dot{a}}}_{GT}=0$, and a temperature difference $\Delta T$ is achieved for the channel relative to the anode and cathode which are held at ambient temperature $T$.  The net change $\Delta T$ is the sum of two contributions $\Delta T_{ind}$ from TI and $\Delta T_J$ from Joule heating.  From here on I analyze specifically the bare channel system illustrated in Fig.~\ref{fig:plank}(b) where heat flows between the channel and the leads only.   Making use of Eqs.~(\ref{gamm112},~\ref{kappa},~\ref{2vardyneq3},~\ref{Jind}) one obtains
\begin{equation}
\Delta T_{ind}= \tau_{GT}^* \gamma e^2 W^2 = \frac{\kappa\tau_{GT}^* GW^2}{C_{GT}} \,,    \label{DeltaTind}
\end{equation}
which describes the effect of TI cooling.  Accounting for Joule heating leads to a fully-balanced  stationary state as $P_{cool}-P_J=-h\left(T_{GT}-T\right)$, or:
\begin{equation}
  T_{GT}-T=  \left( 1+r\kappa    \right)\frac{GW^2}{h}  \,.    \label{TGT}
\end{equation}

One surprising ramification of Eq.~(\ref{DeltaTind}) is the possibility of $\Delta T$ being negative under some conditions.  This may at first seem counterintuitive,  i.e., for one to run current through a conductor and have it get cooler.  Equation~(\ref{DeltaS}) accounts for this possibility and it is clear that only part of the total system has reduced entropy.
Without the DR, having $\Delta S_{GT}<0$ would violate the 2nd law of thermodynamics.  The reservoir concept becomes evident;  The DR produces entropy at such a great rate that subsystems are allowed to lend, or dump, positive amounts of entropy production to the DR.  There are, however, limits to how far this lending can go.  The gate cannot continue to get further and further from equilibrium as the $L_{22}X_{GT}^2$ entropy production term builds up as the magnitude of $X_{GT}$ gets larger.  Eventually, balance is achieved at the stationary state.
Never, during the formation of the stationary state, is the total entropy production rate negative. In fact, induction cooling actually enhances the total rate of entropy production to a more positive value.  This is the physical meaning of Eq.~(\ref{SdotExtra}) which itself constitutes an example of Theorem 2 from Ref.~\cite{Patitsas2014} and can be thought of as a dynamical, irreversible, version of Le Chatelier's principle.  In Le Chatelier's principle, a variable is chosen to be displaced away from equilibrium and other variables relax in such a way as to reduce the magnitude of change in the force conjugate to the chosen variable~\cite{LandauL}.  The change in the system entropy is also reduced in magnitude by the relaxation.  Internal degrees of freedom respond in such a way as to push the system closer towards full equilibrium.
 In the irreversible case, the DR plays the role of the chosen variable.  Other variables, here the gate, respond in such a way as to increase the rate of entropy production by the gate, and to get the entire system to equilibrium faster~\cite{Patitsas2014}.  In fact, under the stationary restriction, ${{\dot{a}}}_{GT}=0$, the total entropy production of the system is maximized. 
Clearly there is more to these stationary states than merely being stationary.  There is also the physical interpretation of the state as one that facilitates, or induces, the approach to equilibrium for the entire system.  In this case, internal degrees of freedom respond in such a way as to push the system faster towards full equilibrium.   


 To further understand these stationary states, or inducer states, it's helpful to look at the probability distribution for the gate variable.  One takes the equilibrium Gaussian function $P_0$ from Eq.~(\ref{P0}) and multiplies by $\exp{\left(\dot{S}_{DR}\tau_{GT}^*/k_B\right)}$.
Using Eq.~(\ref{SdotDR}), this gives
\begin{equation}
P(a_{GT})=c'~\exp{\left(\frac{-a_{GT}^2}{2k_B c_{22}} +\frac{\tau_{GT}^*\gamma X_{DR}^2 a_{GT}}{k_B}\right)}~\, ,   \label{PaGT}
\end{equation}
where $c'$ is a normalization constant.
By inspection, this distribution is skewed and the mean value is readily calculated to give the same result as in Eq.~(\ref{aGTSS}).  The ubiquitous $\tau_{GT}^*$ factor suggests a physical interpretation as a fundamental time step.  After each such time step the DR entropy increases by an amount $\Delta S_{DR}=\dot{S}_{DR}\tau_{GT}^*$ and the number of accessible microstates increases by a factor of $\exp{\left(\Delta S_{DR}/k_B\right)}$.  Sampling all microstates with equal weight is the essential idea behind Eq.~(\ref{expS}), and it is not surprising that the probability  distributions for some variables become skewed.  In the short time step $\tau_{GT}^*$ some values of $a_{GT}$ will facilitate/induce higher levels of entropy production by the DR, and this will give more statistical weight to those $a_{GT}$ values.  It's worth pointing out that this simple time step interpretation seems to hold even though $\tau_{GT}^*$ is below the expected threshold for coarse graining.  It's also worth mentioning that even though $\tau_{GT}^*$ is very small, the factor $\exp{\left(\dot{S}_{DR}\tau_{GT}^*/k_B\right)}$ can be much larger than unity under the right conditions.  It's also helpful to add in some reasoning commonly applied to chemical systems, since it's well known that for a chemical reaction, the $\Delta S$ value for the reaction can play an important role in guiding the outcome.  For this reason these values are measured and tabulated for many reactions.  Here, we apply that idea for each time step, and the $\Delta S$ repeatedly guides and skews the outcome~\cite{Patitsas2014} via Eq.~(\ref{expS}).  Thermodynamic induction has recently been used as an attempt to explain certain STM-based manipulation techniques, for example why Xe atoms can be dragged so effectively across cold metal surfaces~\cite{Patitsas2015}.  The chemical thinking just described seems to apply well to the adsorbate-dragging case.  I believe the same type of thinking is applicable to TI cooling as well. Despite the entropy penalty given by Eq.~(\ref{DeltaS}), the DR conductance is greater and the system overall is allowed to approach equilibrium faster by having the gate induced to lower temperature.
  The time step interpretation also helps to understand Eq.~(\ref{DeltaTind}), since during each time step the DR dissipates an amount of energy equal to $\tau_{GT}^* GW^2$.  A fraction $\kappa$ of that energy results in the induced temperature change.
Finally, since the factor $\exp{\left(\dot{S}_{DR}\tau_{GT}^*/k_B\right)}$ applies to a thermodynamic variable it might also apply to individual microstates for the gate.  If these microstates are labelled $i$ then I postulate the following probability distribution to be applied as a weighting factor for each such microstate:
\begin{equation}
P_i=\frac{1}{Z} e^{\tau_{GT}^*\dot{S}_i/k_B}\, , \label{Pi}
\end{equation}
where $\dot{S}_i$ is the rate of change of the DR entropy when the gate is in state $i$, and $Z$ is a normalization factor.  I propose that this probability distribution could have wide application in analyzing nonequilibrium systems beyond the type of cooling considered here.   This weighting factor would be factored into general thermodynamic calculations much like the Boltzmann factor is in equilibrium studies.  
%

At this point, analysis of specific systems is in order.  As it turns out, analysis of conventional transport at low temperatures leads naturally to analysis of ballistic transport so I will treat the ballistic case first.

\section{Ballistic Transport}\label{sec:ball}

At very low temperatures, carrier scattering becomes less important in determining the electrical conductance, and the boundaries of the conductive channel assume a prominent role~\cite{Datta2005}.  In the absence of impurity scattering, the channel conductance $G=I/W$ is determined by an approach taken by Landauer which gives the electrical current as a summation over all electronic states, with weighting in the summation given by the transmission $\bar{T}$ for each such state, as well as the occupancy for both anode and cathode given by the Fermi-Dirac distribution, $f_0(x)=(1+e^x)^{-1}$~\cite{Datta2005,Landauer1970,Anderson1981,Azbel1984,Landauer1985}.  The Landauer formula specifies the current $I$ in terms of the applied bias $W$ as  
\begin{equation}
I= \frac{2e}{h}\int_{-\infty}^{\infty}dE \, \bar{T}(E)\left[f_0(\beta E-\beta eW)-f_{0}(\beta E)\right]   \,,    \label{Landauer}
\end{equation}
where $\beta=1/k_B T$.  Note that $\bar{T}$ includes summation over subbands arising from the three dimensional nature of the channel.  This function also includes information on band edges.  Even in the case where there is no tunneling barrier to inhibit carrier transmission,  $\bar{T}$ will be zero for energies in band gaps.  When the temperature is low enough, the step-like nature of $f_0$ can be exploited to extract the transmission function as 
\begin{equation}
\bar{T}\approx\frac{h}{2e^2}\frac{\partial I}{\partial W}  \,.    \label{barT}
\end{equation}

The temperature dependence of $I$ and hence $G$ being contained in the two instances of the $f_0$ function allows for straightforward calculation of $\gamma$ using Eq.~(\ref{gamm112}).  If $\bar{T}(E)$ is constant over all energies then by inspection of Eq.~(\ref{Landauer}), the current is independent of temperature.  This implies a complete absence of thermodynamic induction.  This observation has significant bearing on the following analysis.  For systems to exhibit induction effects, the transmission must have structure. The existence of band edges, resonant tunneling, and subbands provide some ways to produce such important structure. 

The dimensionless parameter $\kappa$ is a good measure of the extent of TI.  Using Eq.~(\ref{kappa}), then for fixed applied bias
\begin{equation}
\kappa=\frac{T}{I}\frac{dI}{dT}=-\frac{\beta}{I}\frac{dI}{d\beta}   \, ,
\end{equation}

\begin{equation}
-\beta\frac{\partial I}{\partial \beta}= \frac{2e}{h}\int_{-\infty}^{\infty}dE \,\bar{T}(E) \left[g(\beta E-\beta eW)-g(\beta E)\right] \,,    \label{dIdb}
\end{equation}
where $g(x)=-xf_0'(x)=x[4\cosh^2 (x/2)]^{-1}$.


\subsection{Slowly Varying Transmission Function}

Integration by parts in Eq.~(\ref{dIdb}) gives:
\begin{equation}
-\beta\frac{\partial I}{\partial \beta}= \frac{2e}{h\beta}\int_{-\infty}^{\infty}dE \, \frac{d\bar{T}}{dE} \left[+h(\beta(E-eW))-h(\beta(E)) \right] \,,    \label{dIdbparts}
\end{equation}
where 
\begin{equation}
h(x)=-\frac{x e^x}{1+e^x}+\ln(1+e^x) = \ln(\cosh\frac{x}{2})-\frac{x}{2}\tanh\frac{x}{2} +\ln 2 \, .  \label{F}
\end{equation}
The function $h(x)$ is even and peaked at the origin with inflection points near $x=\pm 1.534$ at which the function falls to 67\% of its peak value.  
For large $x$, $h(x)$ falls off rapidly as $xe^{-x}$.  The area under the curve is $\pi^2/3$.

If $\bar{T}$ varies slowly with energy (compared to $k_B T$) then the $h$ functions in Eq.~(\ref{dIdbparts}) will act like filter functions with the the result:
\begin{equation}
-\beta\frac{\partial I}{\partial \beta}=\frac{2\pi^2 e}{3 h\beta^2}\left[\frac{d\bar{T}}{dE}(eW)-\frac{d\bar{T}}{dE}(0)\right] \,,    \label{dIdbSlow}
\end{equation}
This confirms that $\gamma$ will be small when $\bar{T}$ varies slowly.  A linear term correcting for slowly-changing variations in an otherwise constant transmission will still not result in any TI.  If $W$ is small enough such that the transmission does not vary appreciably over the range 0 to $eW$, then the channel conductance is given by $G=e^2\bar{T}(0)/h$ and the dimensionless TI parameter $\kappa_{ball}$ is given by
\begin{equation}
\kappa_{ball}=\frac{\pi^2 k_B^2 T^2 }{3 \bar{T}(0)}\frac{d^2\bar{T}(0)}{dE^2} \,.    \label{kappaSlow}
\end{equation}

\subsubsection{Low Temperature Limit}

When $k_B T$ is smaller than the energy scale at which the finest structure is found in the transmission, then  the transmission will appear to be slowly varying and Eq.~(\ref{dIdbSlow}) will apply.  Thus, in all systems at very low temperatures, $\kappa_{ball}$ goes as $T^2$ and TI becomes progressively weaker as the temperature is lowered.


\subsection{Rapidly Varying Transmission Function}

To find cases where the effects of TI will be strong in ballistic systems, it makes good sense to look for systems where $\bar{T}$ varies rapidly, for example, as a sharp peak or a sharp step.  

\subsubsection{Sharp Peaks}

Here I first consider the transmission to take the form $\bar{T}(E)=c\delta(E-E_p)$, using the Dirac delta function.  By Eq.~(\ref{Landauer}) the current is given by
\begin{equation}
I= \frac{2e c}{h} \left[f_0(\beta E_p-\beta eW)-f_{0}(\beta E_p)\right]   \,,    \label{LandauerDelta}
\end{equation}
and by Eq.~(\ref{dIdb}) 
\begin{equation}
-\beta\frac{\partial I}{\partial \beta}= \frac{2e c}{h} \left[g(\beta E_p-\beta eW)-g(\beta E_p)\right] \,.    \label{dIdbDelta}
\end{equation}
For the following analysis, the applied bias $W$ is taken as a positive quantity, with no loss in generality.

In the limiting case where $\beta e W<<1$, Eq.~(\ref{LandauerDelta}) gives
\begin{equation}
I= -\frac{2e c}{h}\beta eW f_0'(\beta E_p)   \,,    \label{Landauer2}
\end{equation}
and Eq.~(\ref{dIdbDelta}) gives
\begin{equation}
-\beta\frac{\partial I}{\partial \beta}= -\frac{2e c}{h}\beta eW g'(\beta E_p) \,.    \label{dIdb2}
\end{equation}
The ratio gives
\begin{equation}
\kappa_{ball}= \frac{g'}{f_0'}=-1-\frac{xf_0''}{f_0'}=-1+\beta E_p\tanh(\beta E_p/2) \,.    \label{kappa2}
\end{equation}
Thus $\kappa_{ball}=-1$ is achieved if $E_p=0$, and -1 is the global minimum.

When the sharp peak at $E=0$ is accompanied by another distinct and equally weighted peak at higher energy then one can show that at $eW$ values near this 2nd peak $\kappa_{ball}= g'/(0.5+f_0')$.  The lowest $\kappa_{ball}$ value of $\approx -0.1706$ is achieved at an applied bias $\approx 1.384 ~k_B T/e$ above the peak position.  Finally, for the non-metallic case of just one peak well-removed from $E=0$, $\kappa_{ball}= g/f_0$ and the lowest $\kappa$ value of $\approx -0.2785$ is achieved at an applied bias $\approx 1.2785 ~k_B T/e$ above the peak position. 

To summarize, I have arrived at three general-purpose rules (the -1 rule, the -0.17 rule and the -0.28 rule) for predicting $\kappa_{ball}$ values favourable for TI cooling when the transmission has a sharp peaked structure.

\subsubsection{Sharp Steps}

Here I consider specifically the case where $\bar{T}=1$ at energies below $E_S>0$ above which it equals 2.  From Eq.~(\ref{Landauer}), $I=e^2 W/h$ right up to $W=E_s/e$, while Eq.~(\ref{dIdbparts}) is used to show that $\partial I/\partial\beta$ is very small along the flat section of the step, and that near the step edge
\begin{equation}
-\beta\frac{\partial I}{\partial\beta}=\frac{2e}{h\beta}h(\beta E_s-\beta eW)   \,.  \label{step}
\end{equation}
This gives a strictly positive result $\kappa=(k_B T/E_s) h(\beta E_s-\beta eW)$ that peaks right at $W=E_s/e$ with value $+\ln 2$.  Only  downward steps in $\bar{T}(E)$ give  negative $\kappa$.  Also,  the magnitude of $\kappa_{ball}$ will be small for steps located at energies much larger that $k_B T$, because of the normalization by $I$.

\subsection{General Result}

Summarizing so far, negative and significant $\kappa$ values as low as -1 can be produced when the transmission as a function of energy is highly peaked, whereas, in contrast, a slowly varying transmission will not give significant $\kappa_{ball}$ values.  Here, a variational approach is taken to show that -1 is the lower bound for all possible transmission functions.  
Equation~(\ref{Landauer}) can be written concisely in the form $I= \frac{2e}{h}\int_{-\infty}^{\infty}dE \phi(E) \bar{T} $ and $-\beta\frac{\partial I}{\partial\beta}= \frac{2e}{h}\int_{-\infty}^{\infty}dE \gamma(E) \bar{T} $.  The ratio $\kappa$ is a functional of $\bar{T}(E)$ and the variation $\delta\kappa$ is given by
\begin{equation}
\delta\kappa_{ball}=\frac{\int_{-\infty}^{\infty}dE \gamma \delta\bar{T}}{\int_{-\infty}^{\infty}dE  \phi \bar{T}} -\frac{\int_{-\infty}^{\infty}dE \gamma \bar{T}}{\left[\int_{-\infty}^{\infty}dE  \phi\bar{T}\right]^2}\int_{-\infty}^{\infty}dE \phi \delta\bar{T}  \,.
\end{equation}
Setting $\delta\kappa_{ball}=0$ gives
\begin{equation}
\frac{\int_{-\infty}^{\infty}dE \phi \delta\bar{T}}{\int_{-\infty}^{\infty}dE \phi \bar{T}}=\frac{\int_{-\infty}^{\infty}dE \gamma \delta\bar{T}}{\int_{-\infty}^{\infty}dE \gamma \bar{T}} \, .
\end{equation}
Since $\delta\bar{T}$ is arbitrary
\begin{equation}
 \phi(E) \int_{-\infty}^{\infty}dE' \gamma \bar{T}= \gamma(E)\int_{-\infty}^{\infty}dE' \phi \bar{T}  \, ,
\end{equation}
and since $\phi$ and $\gamma$ are different functions, this condition is impossible to satisfy with any non-zero function except for the delta function $\bar{T}=c\delta(E-E')$.
 
This establishes a proof that I express as a theorem for TI under conditions of ballistic transport:

\vspace{0.25in}
\textit{Theorem:}
\vspace{0.1in}

For nonequilibrium electrical systems governed by ballistic transport,  $\kappa_{ball}\geq-1$.
\vspace{0.2in}

The extremal value $\kappa_{ball}=-1$ is only achieved when the applied bias $W$ is small, and the transmission $\bar{T}(E)$ has the structure of a sharp peak located at $E=0$.

When this theorem is combined with Eq.~(\ref{Jind}) and the physical input that $\tau_{GT}^*<\tau_{GT}$ the following important corollary is established:

\begin{equation}
J_{ind} \geq  -P_J               \, \label{corrol}
\end{equation}
i.e.,

\vspace{0.25in}
\textit{Corollary (no net cooling):}
\vspace{0.1in}

Net cooling is not possible when electrical transport is ballistic.  Thermodynamic induction can provide substantial cooling but it will always be matched or exceeded by Joule heating, at least in the realm exclusive to ballistic transport.
\vspace{0.2in}

\subsection{Subbands}

For a one-dimensional open channel, the transmission $\bar{T}$ in the Landauer formula, Eq.~(\ref{Landauer}), is set to unity, which would seem to give $\kappa_{ball}=0$, by Eq.~(\ref{dIdbSlow}).  Some discussion on this matter is warranted in regards to the band edges. 

\subsubsection{One Dimensional Systems}

If the lateral dimensions of the channel are very small then the channel effectively becomes a chain and subband summation is not required.  Here this condition is referred to as subband inactive.  While technically a high enough bias can always be applied to access subbands no matter how thin the chain is, practically speaking such conditions are difficult to access when the chain thickness is right down to atomic dimensions. In this subband inactive case the current can be given in terms of a state sum as 
\begin{equation}
I_{1d} = \frac{2e}{h}\frac{2\pi}{L_x}\sum_{k_x}^{}  \, \bar{T}_{1d}(E)\frac{\partial E}{\partial k_x}\left[f_0(\beta(E-eW))-f_{0}(\beta E) \right]  \,,    \label{I1Dkx}
\end{equation}
where $L_x=l$ is the length of the channel.     The transmission $\bar{T}_{1D}$ is actually the transmission coefficient and would equal unity for an open channel.  Also implied here is a dispersion relation $E(k_x)$ which includes lower and upper band edges.  The physical significance of $\frac{\partial E}{\partial k_x}$ in Eq.~(\ref{I1Dkx}) is the group velocity along the channel, and it must be stressed that proper use of this only occurs in the continuum limit. 

The summation $\sum_{k_x}^{}$ can be replaced by $\frac{L_x}{2\pi}\int dk_x$ if the channel length is much longer than the size of the primitive unit cell $a$, a condition referred to here as an extended channel.  When this integral is converted to one over energy, the group velocity factor is exactly canceled by a density-of-states factor.  If $E_1$ ($E_2$) denotes the lower (upper) energy band edge, then for an open channel
\begin{equation}
I_{1d}= \frac{2e}{h}\int_{E_1}^{E_2}dE \, \left[f_0(\beta(E-eW))-f_{0}(\beta E) \right]  \,.    \label{I1}
\end{equation}
where we note the key difference to Eq.~(\ref{Landauer}) is the existence of the limits at $E_1$ and $E_2$.
Making use of Eq.~(\ref{dIdbparts}) I obtain
\begin{equation}
\frac{\partial I_{1d}}{\partial \beta}= \frac{2e}{h\beta} \, \left[h(\beta(E_1-eW))-h(\beta(E_2-eW)) -h(\beta E_1)+h(\beta E_2) \right]  \,.    \label{dI1b}
\end{equation}


This exact result works well for wide bands, so with band edge values much larger in magnitude than both $k_B T$ and $eW$, $\gamma$ will be very small, and hence TI effects will be very small.
This result holds even if there is considerable structure in the $E(k_x)$ function within the band edges, as mentioned, a result of the group velocity canceling the density of states function.  This result holds as long as the density of states is a continuous function, i.e. for the extended channel.  If the channel becomes short enough then the density of states is a series of discrete peaks, and the analysis presented earlier takes over.  The important parameter to decide if a channel density of states is continuous or discrete is the so-called coupling energy.  This parameter $\gamma_c$ describes the amount of coupling between a given discrete state of the channel and the leads (anode and cathode), and may be evaluated by calculation of the self-energy component of the single-particle Green's function of the contact (leads) coupled to the channel.  Calculation of this coupling would follow the nonequilibrium Green's function formalism~\cite{Schwinger1959,Baym1962,Keldysh1965}.  If the coupling energy exceeds the typical energy spacing between peaks then the channel can be considered as extended, and the density of states function as continuous.

\begin{figure}[ht]
	\centering
		\includegraphics[width=0.9\textwidth]{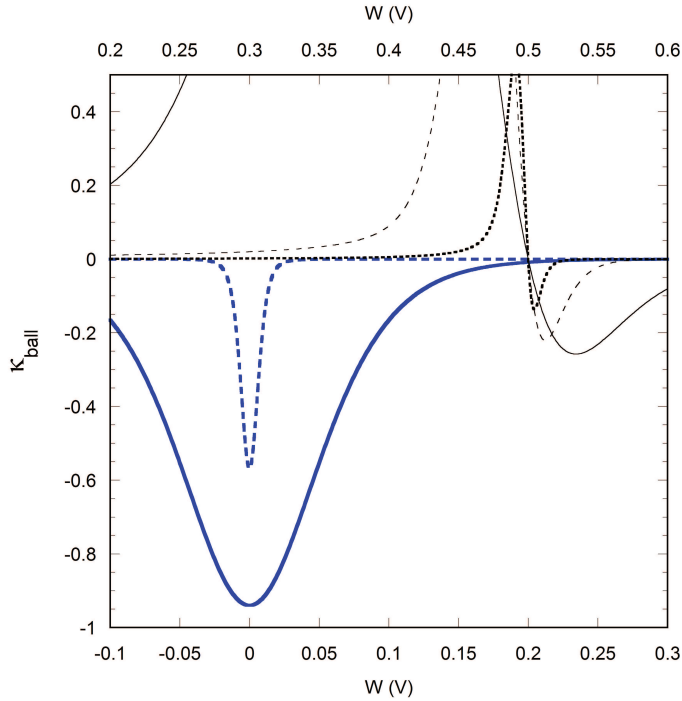}
\caption{\label{fig:restunn} Plots of $\kappa_{ball}$ vs. applied bias for a resonant tunneling channel. For a resonance at 0 eV plots are shown at 300 K (thick, blue, solid) and 30 K (thick, blue, dashed), and for resonance at 0.5 eV plots are shown at 300 K (thin, black, solid), 100 K (thin, black, dashed), and 30 K (thin, black, dot-dash).  }
\end{figure}

A similar discussion applies to an extended channel containing a resonant tunneling barrier.  If the width in energy of the resonance is less than $k_B T$ then the transmission will be considered to have a single sharp peak and values for $\kappa_{ball}$ near -0.28 or -1 are possible, depending on the position of the peak.  Figure~\ref{fig:restunn} shows $\kappa$ plotted as a function of applied bias for an extended one-dimensional system with with band edges at $\pm 1.0$ eV.  The resonances is described by a transmission given by 
\begin{equation}
\bar{T}=\frac{\gamma_1 \gamma_2}{(E-E_r)^2+\bar{\gamma}^2}
\end{equation}
where $\bar{\gamma}=(\gamma_1+\gamma_2)/2$~\cite{Datta2005}.  In this calculation the two coupling parameters to each lead are taken to have the same value of 3.0 meV.  In Fig.~\ref{fig:restunn} plots are made for two values of $E_r$, 0.5 eV and 0 eV.  At 300 K temperature, the thick (blue) solid curve is symmetric about $E=0$ and dips down to -0.94, which agrees well with the -1 rule.  At the lower temperature of 30 K, the transmission no longer appears as sharp compared to $k_B T$ and the extent of the dip is reduced to -0.57.  Also the width of the dip is much smaller in the thick (blue) dashed curve.  Three calculations were made for a resonance at +0.5 eV at 300 K (thin solid), 100 K (thin dashed), and 300 K (thin dash-dot), as shown in Fig.~\ref{fig:restunn}.  The 300 K curve dips down to a minimum value of -0.26, near the expectation of the -0.28 rule.  Also, as expected the minimum does not occur at $E_r/e$ but rather at higher bias.  The 100 K and 30 K plots show this shift is proportional to the temperature, as expected.

In summary, significant TI is possible in 1-dimensional systems that are either very short open channels, or have a resonant tunneling structure.

\subsubsection{Two and Three Dimensional Systems}

For an open 2-dimensional channel, Eq.~(\ref{I1Dkx}) is generalized to
\begin{equation}
I_{2d} = \frac{2e}{h}\frac{2\pi}{L_x}\sum_{k_x}^{} \sum_{k_y}^{} \, \frac{\partial E}{\partial k_x}\left[f_0(\beta(E-eW))-f_{0}(\beta E) \right]  \,.    \label{I2Dkx}
\end{equation}
For the extended case
\begin{equation}
I_{2d} = \frac{L_y}{2\pi}\frac{2e}{h}\int_{}^{} \int_{}^{}dk_x dk_y \, \frac{\partial E}{\partial k_x}\left[f_0(\beta(E-eW))-f_{0}(\beta E) \right]  \,.    \label{I2Dkxb}
\end{equation}
This double integral can be converted to an integral over energy alone by considering the integral over $k_x$ and $k_y$ over a thin region of k-space that includes only states with energies between $E$ and $E+dE$.  The thickness of this ribbon in the $x$ direction is $dE/|\frac{\partial E}{\partial k_x}|$ so again there is exact cancellation of the x-component of the group velocity.  Integration over $k_y$ gives the parameter $\Xi(E)$ equal to the maximum extent of the shell in the $y$ direction, leaving 
\begin{equation}
I_{2d} = \frac{L_y}{2\pi}\frac{e}{h}\int_{}^{} dE \, \Xi(E)\left[f_0(\beta(E-eW))-f_{0}(\beta E) \right]  \,.    \label{I2Dkxc}
\end{equation} 
For many of the allowed energy values in the band, one would expect $\Xi(E)$ to be on the order of $2\pi/a$ where $a$ is the typical unit cell dimension on the nanometer scale.  We see that $\frac{L_y}{2\pi}\Xi$ and $\bar{T}$ are the same function for an open channel and on the order of $L_y/a$ which is roughly the number of subbands expected for an open channel.  The same conclusion about subband number holds true for a 3-dimensional channel.  Thus, for an open channel,  $\bar{T}$ is the integrated density of states function $N_{d-1}(E)$ corresponding to the directions transverse to the current flow.  Since the transverse density of states is the derivative of this function, i.e., $D_{d-1}=dN_{d-1}/dE$, making use of Eq.~(\ref{dIdbparts}) gives
\begin{equation}
-\beta\frac{\partial I}{\partial \beta}= \frac{2e}{h\beta}\int_{-\infty}^{\infty}dE \, D_{d-1}(E) \left[+h(\beta(E-eW))-h(\beta(E)) \right] \,.    \label{dIdbDOS}
\end{equation}
The subscript $d-1$ is a reminder that for a 3-dimensional channel, the transverse density of states resembles that of a 2-dimensional system, etc.  Even though structure in the density of states in the flow direction is canceled out, such structure in the transverse directions does come through to affect TI through the $D_{d-1}$ function.  I point out that since $N_{d-1}$ might sometimes decrease with energy, $D_{d-1}$ is not positive definite.
For temperatures low enough such that $D_{d-1}$ does not vary much over the energy interval $k_B T$, the localized nature of $h(x)$ can be exploited to give:
\begin{equation}
-\beta\frac{\partial I}{\partial \beta}= \frac{2e}{h\beta^2}\frac{\pi^2}{3} \left[  D_{d-1}(eW) - D_{d-1}(0) \right] \,.    \label{dIdbDOS2}
\end{equation}
If the channel becomes subband active by having the lateral dimension increased, one expects a simple staircase structure for $N_{d-1}$.  From Eq.~(\ref{dIdbDOS2}) it's apparent that an upwards step (while $W$ is positive) will always give a positive spike in $\kappa_{ball}$ as $eW$ is scanned through the step energy.  Negative $\kappa_{ball}$ will only be found for downward steps in $N_{d-1}$.

For the case where the channel is extended in all three directions, the function $D_2$ plays the important role and it is known to be constant when the dispersion relation is parabolic.  In this case Eq.~(\ref{dIdbDOS2}) shows that there would be no TI heat transfer.  In many systems the dispersion relation is not parabolic over the entire band, so there may be some 3-dimensional fully extended systems showing appreciable ballistic TI heating or cooling.
For the case where the channel is extended in the $x$ direction and one transverse direction, but subband inactive in the other transverse direction, it is the one-dimensional function $D_1$ which enters Eq.~(\ref{dIdbDOS2}).  In this case $D_1(E)$ goes as $E^{-1/2}$, again for parabolic dispersion, with the result that only TI heating is possible; $\kappa_{ball}$ will always be positive. 
Thus in all channels with dimensions either extended or subband inactive it seems the possibilities of observing TI cooling are small.

If however one, or both, of the transverse dimensions is subband active then sharp peaks in $D_{d-1}$ are expected and near these peaks significant and negative $\kappa_{ball}$ values are expected;  These structures may be described as ribbons and thick sheets. 

\begin{figure}[ht]
	\centering
		\includegraphics[width=0.9\textwidth]{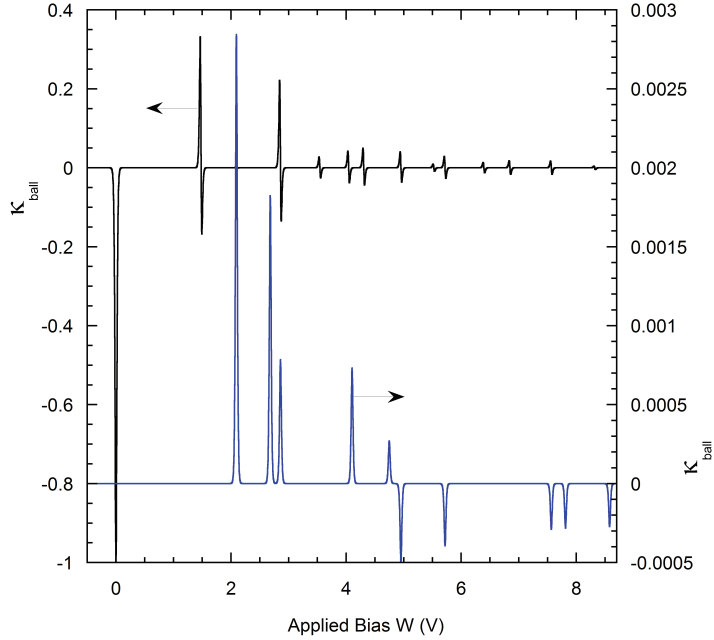}
\caption{\label{fig:zigm6} Plots of $\kappa$ vs. applied bias for ballistic conduction in zigzag carbon nanotubes.  Numerical computation was performed for the case $m=6$ (0.93 nm diameter), with short (5.04 nm length, left axis, black curve) and long (84 $\mu$m, right axis, blue plot) nanotubes.  Both calculations were done at 100 K temperature. }
\end{figure}

\subsection{Numerical Calculations for Carbon Nanotubes}

Many of these results presented in this section are well-illustrated by specific numerical calculations performed on carbon nanotubes.
The carbon nanotube is formed by rolling up a sheet of graphene layer with carbon atoms arranged in a hexagonal pattern sheet with bond length $a_0=0.28$ nm, primitive unit cell vectors $\overrightarrow{a_1}=a\hat{x}+b\hat{y}$ and $\overrightarrow{a_2}=a\hat{x}-b\hat{y}$, where $\hat{x}$ runs along the tube axis, $a=0.42$ nm, and $b=0.24$ nm.  The Brillouin zone is hexagonal-shaped with vertices at $(\pm\pi/a,\pm\pi/3b)$ and $(0,\pm 2\pi/3b)$.  A simple tight-binding model, well-described in Ref.~\cite{Datta2005}, is implemented with a dispersion relation given by
\begin{equation}
E(k_x,k_y)=\pm t\sqrt{1+4\cos k_y b\cos k_x a +4\cos^2 k_y b} \,,  \label{disp}
\end{equation}
where $t=1.43$ eV.  States with $E=0$ are found at the vertices $(\pm\pi/a,\pm\pi/3b)$.  When the sheet is rolled up to form a zigzag nanotube, periodic boundary conditions confine $k_y$ to discrete lines parallel to the x-axis, i.e., $k_y=\pi\nu_y /mb$ where $\nu_y$ is a running integer index and the integer $m$ determines the nanotube diameter as $d = 2bm/\pi$.  If $m$ is an integer multiple of 3, then the $E=0$ states are accessed and the nanotube channel will be metallic.  Equation~(\ref{I2Dkx}) is used to numerically sum up for the current as a function of applied bias and temperature.  The temperature derivative is evaluated numerically to produce $\kappa_{ball}$.  The calculation leaves out electron-electron interactions: effects from possible charge buildup in this inflow/outflow problem~\cite{Datta2005}, and is simply intended to emphasize the effects of band and subband structure on $\partial I/\partial T$.  

\begin{figure}[ht]
	\centering
		\includegraphics[width=0.8\textwidth]{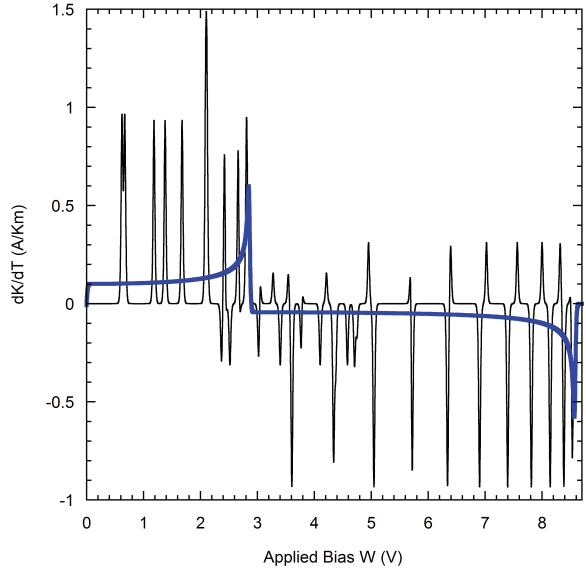}
\caption{\label{fig:zigbigm} Plots of temperature derivative of current density vs. applied bias ($\partial K/\partial T$ vs. W) for ballistic conduction calculations with $m=24$ (3.7 nm diameter, left axis, thin black curve) and with $m=1601$ (247 nm diameter, right axis, thick blue curve).  In both cases the zigzag carbon nanotube length is 84 $\mu$m long and the temperature is 100 K.}
\end{figure}

Results for $\kappa_{ball}$ as a function of applied bias for a rather small zigzag nanotube (0.93 nm diameter and 5.04 nm length) at 100 K temperature is shown in Fig.~\ref{fig:zigm6}.  A prominent downwards peak (dip) at zero bias goes almost down to $\kappa=-1$ just as the -1 rule predicts.  The existence of such a peak is consistent with the channel being metallic.  Above 0.25 V, $\kappa_{ball}$ becomes very small until a peak-dip pair is encountered near 1.5 V and as the -0.17 rule predicts, a positive peak is quickly followed by a second peak that indeed dips down to $\approx -0.17$.  Successive peaks are also found and these fall off in magnitude as the electrical current increases.  Overall this plot is consistent with a molecular type system with discrete and well-spaced energy levels.
Though such short chains are difficult to produce and connect into electrical circuits, STM-based techniques are known to produce very small wires made of gold which span the space between two STM probes~\cite{Lou2010}. The STM probes play the role of the anode and cathode depicted in Fig.~\ref{fig:plank}(b).  The length of these wires can be very short, from approximately 100 nm down to 10 nm or perhaps even shorter.  These specialized gold wires may have electronic structure resembling a series of discrete peaks and hence $\kappa$ vs. bias curves that might resemble the 5.04 nm nanowire calculation.
 
As the channel length is increased in the computations, more peaks are produced as expected until the length exceeds about 200 nm.  For longer nanotubes the peaks coalesce and dramatically weaken, including the $E=0$ peak.  At the temperature of $100$ K, nanotubes longer than 200 nm may be considered as extended in the flow direction.  
Also in Fig.~\ref{fig:zigm6}, a $\kappa_{ball}$ vs. $W$ plot is shown for a zigzag nanotube 84 $\mu$m in length.  The zero energy peak has completely disappeared.  The only structure left is 10 peaks all corresponding to subband structure, i.e, whenever the subband peak number changes either upwards or downwards.  Most striking is the change in the scale in Fig.~\ref{fig:zigm6} with the strongest peak less than 0.003.  Below 4.8 V applied bias only positive $\kappa_{ball}$ peaks are observed.  Only above this where the subband number decreases as the top of the band is approached are negative value peaks observed where TI cooling may occur.  The lack of structure in the extended channel plot is noteworthy;  All band structure related to the $x$ direction is indeed filtered out by the cancellation of the group velocity as discussed above.

Further calculations are presented in Fig.~\ref{fig:zigbigm} in which the effect of increasing the nanotube diameter is explored.  In these plots the derivative with respect to temperature is taken of the sheet current density $K$ calculated as the current divided by the tube circumference.   With an intermediate diameter nanotube, (m=24, 3.7 nm diameter) and 34 $\mu$m long, the plot shows a series of discrete, positive peaks up to a bias of 3 V, after which the peaks point downwards.  Each peak signifies the beginning or end of a subband.  In light of Eq.~(\ref{dIdbDOS2}), given that there is one active dimension transverse to the direction of current flow, it makes good sense to see the density of states function $D_1$ as a series of equally-weighted spikes, consistent with Eq.~(\ref{step}).  This is indeed the case, barring degeneracies, for example the first four peaks from 0.6 V to 1.7 V.

When $m$ is increased to a value of 2001, producing a tube with diameter 309 nm, the subband structure coalesces as demonstrated by the continuous nature of the thick blue curve of  Fig.~\ref{fig:zigbigm}.
Given that this system is a large graphene sheet rolled into a tube it should produce the same results as a graphene sheet extended in both dimensions along the sheet.  If electrical probes are connected to large pieces of graphene, an applied bias of greater than 3 V would be required to produce TI cooling;  For smaller bias levels TI heating would only add to Joule heating.  The amount of cooling at $W=3.5$ V is very small, where $P_J$ would only cancel off 0.0016\% of the Joule heating.  It appears that it will not be easy to observe TI cooling in extended carbon nanotubes and graphene.  In practice it appears to be more likely to observe TI cooling in very small carbon structures, in particular very short along the flow direction.

\subsection{Interpolation with Conventional Transport}

For an open channel under low applied bias, the ballistic conductance is given by
\begin{equation}
G_{ball}=\frac{2e^2 M}{h} \, ,
\end{equation}
where $M=D_{d-1}$ is the subband number and is the same as $\bar{T}$.  Since longer channels will realistically exhibit carrier scattering some analysis for the interpolation between purely ballistic and purely conventional transport is warranted.  The proper interpolation formula is 
\begin{equation}
\frac{1}{G_{net}}=\frac{1}{G_{ball}}+\frac{1}{G_{conv}}=\frac{h}{2e^2 M}+\frac{h}{2e^2 M}\frac{l}{\Lambda} \, ,  \label{interp}
\end{equation}
where $\Lambda$ is a length scale on the order of the mean free path~\cite{Datta2005}.  The second term in Eq.~(\ref{interp}) must match the conventional conductance for a channel $G_{conv}=\sigma A/l$, where $\sigma=ne^2\tau_{ep}/m$ is the electrical conductivity and $\tau_{ep}^{-1}$ is the electron phonon scattering rate.  One would expect $M$ to scale with the channel cross-sectional area $A$, and indeed the correct matching occurs if $M\approx k_F^2 A $ where $k_F$ is the Fermi wavenumber.  For conventional transport, TI will be connected to how the electrical conductivity varies with temperature.  Noting that $\sigma$ is proportional to $\Lambda$, and defining $\kappa_{0}=d\ln\sigma/d\ln T$, the following interpolation formula is obtained for $\kappa$: 
\begin{equation}
\kappa_{net}=\kappa_{ball}+\kappa_{0}\left(1+\frac{l_{mfp}}{l}\right)^{-1} =\kappa_{ball}+\kappa_{conv} \, ,  \label{kapinterp}
\end{equation}
where I have replaced $\Lambda$ with the mean free path defined from $l_{mfp}=mv_F\sigma/ne^2$.  Also, $\kappa_{conv}$ has been defined as the product of $\kappa_0$ and $\left(1+\frac{l_{mfp}}{l}\right)^{-1}$.  In the limit $l<<l_{mfp}$, the reasonable result $\kappa_{net}=\kappa_{ball}$ is obtained.  More interestingly, in the limit $l>>l_{mfp}$ where conventional conductance dominates, $\kappa_{net}=\kappa_{ball}+\kappa_{0}$, i.e., the ballistic contribution persists in a surprising way.  This result has nothing to do with TI.  If a bulk material with a small temperature dependence for the conductivity is studied carefully, the ballistic $\kappa$ specified in Eq.~(\ref{kappaSlow}) may be significant and measurable, particularly at higher temperatures.
As far as the low temperature behaviour goes, if the Bloch-Gr{\"u}neisen law holds and the conventional resistivity goes as $T^5$, then $\kappa_0=-5$ and $\kappa_{conv}$ goes as $T^5$, thus falling off much faster than $\kappa_{ball}$.

The interpolation formula, Eq.~(\ref{kapinterp}), plays an important role in the following discussion on conventional transport.  In many of these systems, $r$ is small and in the quest for systems with larger $r$ values, channels with mean free paths near the channel length will be explored. In such systems the $\left(1+\frac{l_{mfp}}{l}\right)^{-1}$ factor plays an important role.

\section{Conventional Electrical Transport}\label{sec:conv}

In this study I focus on systems with negative values of $\gamma$ and $\kappa$, since positive values will only result in induction heating adding to the inherent Joule heating.  For example, intrinsic semiconductors are a poor choice for cooling even though they are expected to have a large value for $\kappa$, ideally near $E_g/k_B T$ where $E_g$ is the band gap.

\begin{figure}[ht]
	\centering
		\includegraphics[width=0.9\textwidth]{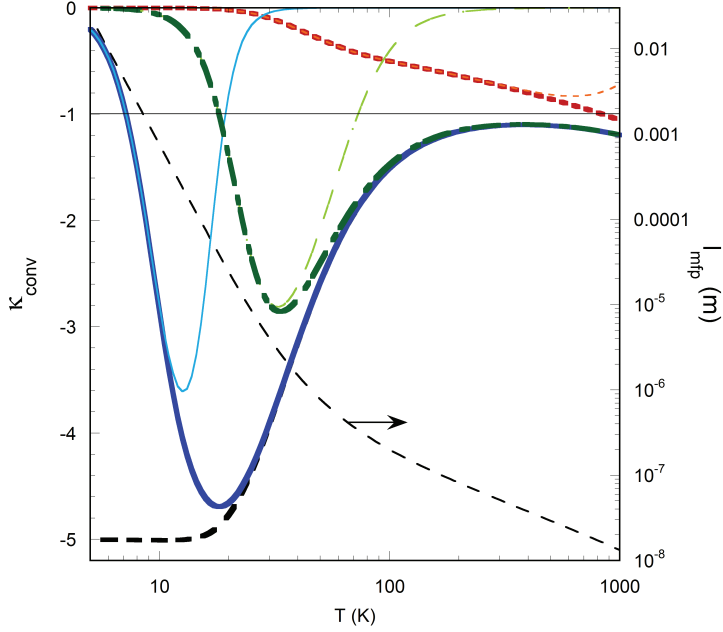}
\caption{\label{fig:silver} For highly purified silver, plots of the logarithmic derivative $\kappa_0=d\ln\sigma/d\ln T$ (thick, black, long-dash curve) and $\kappa_{conv}$ for 1 mm (thick, blue, solid), 10 $\mu$m (thick, green, dash-dot), and 100 nm (thick, red, short-dash) channel lengths, all on the left axis.  Also displayed as thin curves are $r\times\kappa_{conv}$ for the same three channel lengths (1 mm: thin, blue, solid, 10 $\mu$m: thin, green, dash-dot, 100 nm: thin, red, short-dash). Plotted on the right axis is the electron mean free path (thin, black, dashed curve). }
\end{figure}

Conventional metals are good candidates for cooling the channel by induction since their electrical conductivity decreases with temperature.
In Fig.~\ref{fig:silver} calculations are presented that can be used to assess the level of inductive cooling in a conductive channel made of highly purified silver.  Silver is a good choice of material as it is been extensively studied~\cite{Ehrlich1974,Caplin1975,Rumbo1976,Kos1990} and good low temperature results are available after Fe impurities are removed by oxygenation~\cite{Barnard1982}.  In my calculations, the electrical conductivity function was taken from Ref.~\cite{Matula1979} and the derivative with respect to temperature was evaluated numerically.  Matula's $G_1(T)$ function fit for the Bloch-Gr{\"u}neisen function and supplied parameters were used to calculate $\sigma(T)$.  From this function both $\gamma$ and $\kappa$ were calculated using Eqs.~(\ref{gamm112},~\ref{kappa}).  For the specific heat, which also enters into Eq.~(\ref{Jind}), the Debye model with Debye temperature of 215 K was implemented to calculate the lattice contribution to $c_V$~\cite{AM}.  The electronic contribution to the specific heat was also included~\cite{AM}.  In order to calculate the thermal conductivity a value of $2.44\times 10^{-8}$ V$^2$/K$^2$ was used for the Lorenz number~\cite{AM}.  

After producing an accurate calculation for the electrical conductivity vs. temperature over the range from about 1 K up to temperatures well above room temperature, the logarithmic derivative $\kappa_0$ is readily calculated.  This function, which is depicted as the thick (black) short-dashed curve in Fig.~\ref{fig:silver}, stays below -1 throughout the entire temperature range shown.  Below the Debye temperature of 215 K the phonon density as well the electron-phonon scattering rate drop quickly.  Indeed, below $\approx 100$ K, $\kappa_0$ drops down and eventually levels off at the value of -5 as expected from the Bloch-Gr{\"u}neisen law~\cite{Matula1979}.  It should be pointed out that for the case of silver under 10 K, Matthiessen's rule seems to break down and a $\kappa_0$ value near -4 may be more correct~\cite{Barnard1982}.  I will return to this low temperature behaviour below.

	
Before obtaining $\kappa_{net}$, the interpolation formula, Eq.~(\ref{kapinterp}), and hence $\kappa_{conv}$ is required which in turn depends on the channel length $l$.  To illustrate this sample geometry effect, a few representative plots are displayed Fig.~\ref{fig:silver} as the thick curves.  The thick (green) dot-dash curve was calculated as $\kappa_{conv}$ for a sample length of 10 $\mu$m.  At high temperatures the mean free path is much less than $l$ and, as expected, there is little change from the $\kappa_0$ curve. Below $\approx 35$ K the $\kappa_{conv}$ curve bends upwards sharply and at 24 K the mean free path matches the sample length.  At temperatures well below this TI is effectively quenched, at least for conventional transport, as scattering events (and fluctuations) become rare.  The minimum $\kappa_{conv}$ value of -2.85 suggests that a highly purified silver channel of this length would be a good candidate for exhibiting TI cooling near 34 K.  For a longer channel (1 mm), the overall story is similar, with the thick (blue) solid curve dipping down to even lower values, as low as -4.7 at 18 K.  For very small channel lengths TI is manifested at more modest levels and is more significant at higher temperatures, as exemplified by the thick (red) short-dash curve for a 100 nm long channel.

Alone, the dimensionless quantity $\kappa_{conv}$ is a good indication of the strength of TI but knowledge of the cooling power will better allow for quantitative analysis. 
Equations~(\ref{Jind},~\ref{TGT}) can be used to quantify the strength of TI cooling in terms of the Joule heating, if the  ratio $r=\tau^*_{GT}/\tau_{GT}$ is well-characterized.  The ratio should always be smaller than unity since the relaxation time is expected to exceed the timescale for fluctuations that produce the relaxation.  Also, for large systems $r$ is expected to be small.  In metals such as silver, electrons are primarily responsible for thermal conduction so the fluctuation timescale must be the electron-phonon scattering time, i.e., $\tau_{GT}^*=\tau_{ep}=l_{mfp}/v_F$.   Using the Wiedemann-Franz law, $\lambda=L_0 T\sigma$, where here I take $L_0=\pi^2 k_B^2 /3e^2$, then in the limit where the channel length greatly exceeds the mean free path:
\begin{equation}
r=\left(\frac{L_0 n e^2 T }{2 c_V E_F }\right)\frac{l_{mfp}^2}{L_x^2} \,,~~~~~~~l_{mfp}<<L_x \,.  \label{r0}
\end{equation}
It is noted that the first factor in brackets in Eq.~(\ref{r0}) is close to 1/3 at very low temperatures where the specific heat is dominated by electrons, and it never exceeds this value.  When the channel length is less than the mean free path, Eq.~(\ref{r0}) must be modified;  In order to keep $r<1$, the following formula has the appropriate large and small sample limits:
\begin{equation}
r=\left(1+\frac{2 c_V E_F L_x^2}{L_0 n T l_{mfp}^2}\right)^{-1} \, .  \label{r}
\end{equation}
With $P_{cool}=-r\kappa_{net}P_J$ and ignoring ballistic effects, the ratio $P_{cool}/P_J$ is readily calculated, and is presented as the thin curves in Fig.~\ref{fig:silver}.  
\begin{figure}[ht]
	\centering
		\includegraphics[width=0.9\textwidth]{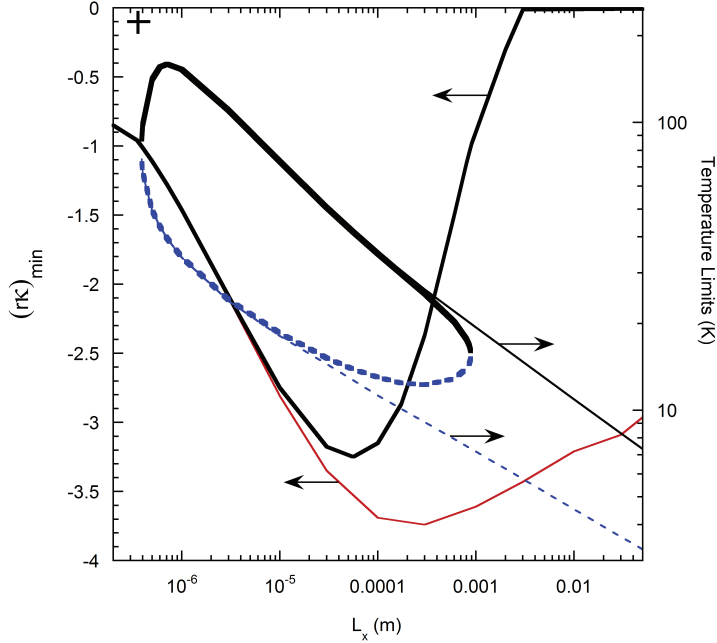}
\caption{\label{fig:limits}  Minimum values of the power ratio $r\kappa_{conv}=J_{ind}/P_J$ versus channel length for the case where the Bloch-Gr{\"u}neisen law holds (thick, black, solid curve on left axis) and the case where a constant resistivity $\rho_0=8\times 10 ^{-12}~\Omega$m was added into the model (thin, red, solid curve on left axis).  On the right axis is plotted upper $T_U$, and lower $T_L$, temperature limits for net cooling, again for the the case where  the Bloch-Gr{\"u}neisen law holds (thin, black, solid and thin, blue, dashed curves), and the case with the same offset resistivity (thick, black, solid and thick, blue, dashed curves).   }
\end{figure}	
For the 10 $\mu$m channel, the thin (green) dash-dot curve represents $r\kappa_{conv}$.  This $r\kappa$ curve lies below -1 in the temperature range 18-73 K and net cooling is expected to be possible in this range (see Eq.~(\ref{TGT})).  In general, the part of the $r\kappa$ curve below -1 represents the window of opportunity for net cooling to occur.  
Observation of net cooling in a rather ordinary electrical circuit would indeed be remarkable.  A clear and simple test for net cooling could involve a conductance experiment where the applied bias is ramped up slowly.  At high enough bias levels one would look for decreases in the sample temperature, or for anomalous increases in sample conductance.  For the longer 1 mm channel, the thin (blue) solid curve shows that net TI cooling will be possible in the somewhat narrower temperature range 7 K to 20 K, though more will be said on this below.
For channel lengths below about 500 nm, the $r\kappa_{conv}$ curve does not dip below -1 and net cooling by conventional transport will not be possible (see thin (red) dashed curve, for example).
For such short channels, net cooling might be marginally possible if ballistic transport is favourable with $\kappa_{ball}\approx -1$ and then with $\kappa_{conv}$ added in, $\kappa_{net}$ might be less than -1.  Also, even if $\kappa_{net}$ is not less than -1, tests for the TI effect are still possible.  More modest levels of TI cooling (or even heating) might be discerned after careful modeling of experiments made on such a system which measure $\Delta T$ vs. applied bias $W$.  If modeling for temperature rises due to Joule heating alone are accurate enough then discrepancies with experimental results might be explained by TI.  It should be mentioned that the temperatures reported here come from a discrete model presented here for TI and must therefore be interpreted as an average over the entire channel.  Though the center of the channel could certainly be somewhat cooler than near the leads, for example, working with the mean value suffices to illustrate the important physics of the TI cooling that might be occurring in such channeled conduction.  

Continuing with the assumption of validity of the Bloch-Gr{\"u}neisen law, in Fig.~\ref{fig:limits}, I show a plot of the minimum value of the power ratio $r\kappa_{conv}=-P_{cool}/P_J$ as a function of channel length, as the thin, red, solid curve.   Values below -1 are sustained over a wide range all the way down to 390 nm.  Net cooling for shorter samples fails essentially because the mean free path is too long which makes $r$ small.  This does not seem to be a fundamental barrier and for materials other than silver this barrier may not even exist.
 For channels longer than 390 nm, $P_{cool}$ will exceed $P_J$ over some range of temperature (limits $T_U$ and $T_L$ defined by $r\kappa_{conv}=-1$ and displayed as thin curves on right axis), for example from $T_L=3.3$ K to $T_U=7.3$ K for a 5 cm long channel.  The lowest value of -3.73 for $r\kappa_{conv}$ is found for a 300 $\mu$m channel. 


\begin{figure}[ht]
	\centering
		\includegraphics[width=0.9\textwidth]{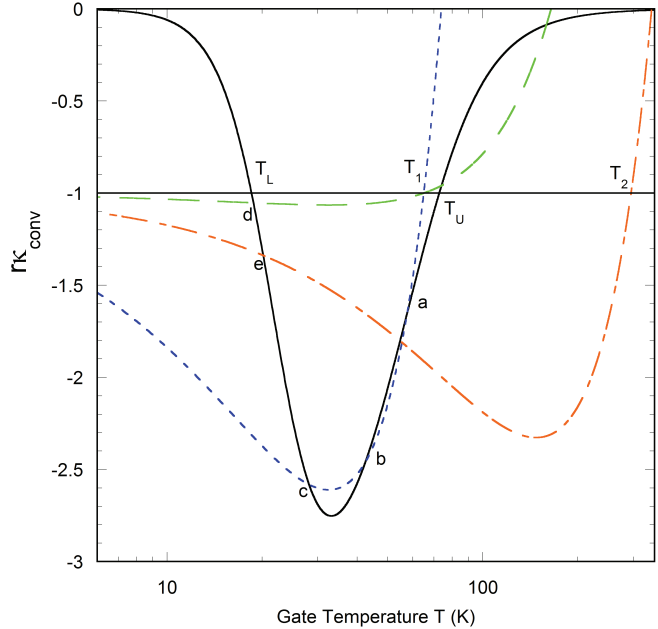}
\caption{\label{fig:ambient} Power ratio $r\kappa_{conv}=J_{ind}/P_J$ versus temperature for a 10 $\mu$m long highly purified silver channel (solid, black curve), as well as plots of the right-hand side of Eq.~(\ref{graph}) for ambient temperature 65 K and $W=4$ mV (blue, short-dash), 65 K and 20 mV (green, long-dash), and 295 K, 20 mV (red, dash-dot). Solutions for the stationary state are found at intersection points with $r\kappa_{conv}$.  These intersections are labelled a, b, c, d, e.}
\end{figure}


The results discussed so far have not incorporated impurity scattering and thus overestimate the relative strength of TI cooling at low temperatures.  In practice, the Bloch-Gr{\"u}neisen $T^5$ law breaks down as impurity and surface scattering become important when phonon scattering disappears. 
The conductivity for highly purified silver samples has been found to flatten below about 3 K to values around $10^{11}$ S~\cite{Kos1990,Barnard1982}. With a small constant added to the Bloch-Gr{\"u}neisen resistivity, the logarithmic derivative $\kappa_{conv}$ goes as $T^{5}$, causing induction cooling to fall off at low temperatures. 
 Incorporating this extra resistance into my numerical model is straightforward and the results are also plotted in Fig.~\ref{fig:limits}.  For channel lengths below about 10 $\mu$m there is no discernible change, but above 100 $\mu$m there is a large detrimental effect, $r \kappa_{conv}$ is cut down considerably, and net cooling is cut off completely for samples longer than 0.9 mm.  Being limited to purified samples no longer than about 1 mm is a serious challenge for actual testing, as samples used in conductance work have been over 10 cm in length~\cite{Kos1990}.  Nevertheless, producing highly purified silver samples shorter than 1 mm is certainly possible and for such samples the experiment appears to be straightforward.  For example, if samples 100 $\mu$m long could be manufactured, the workable temperature range is $T_L=12$ K to $T_U=35$ K.  
The widest temperature range of 120 K is predicted to occur with 700 nm samples.  This sample length also gives the highest possible upper temperature limit for net cooling of $T_U=159$ K.  Similar numerical calculations have also been performed on channels composed of highly purified copper, again using Matula's fit parameters.  For copper, I find the greatest upper temperature limit to be $T_U=222$ K for a 360 nm long channel, displayed as the cross in the upper left of Fig.~\ref{fig:limits}.  This change in $T_U$ for copper is easily explained by noting that the electrical conductivity in this temperature range is determined by the phonon number density, and at temperatures below the Debye temperature the conductivity changes rapidly.  This is the reason for $\kappa_0$ values in the range -2 to -3 at temperatures near 100 K to 200 K.  The higher Debye temperature of 315 K for copper as compared to 215 K for silver explains why $T_U$ is larger for copper.

The extent of TI cooling can be better understood by re-expressing Eq.~(\ref{TGT}) in terms of the electrical conductivity and the Lorenz constant, as:
\begin{equation}
 r\kappa_{conv}= -1+\frac{L_0 T_{GT}}{W^2}  \left( T_{GT}-T\right)  \,.    \label{graph}
\end{equation}
This produces a simple graphical method to understand how the gate temperature behaves relative to the ambient temperature $T$ in terms of applied bias.  As long as $L_0$ remains constant, the right side of Eq.~(\ref{graph}) is a parabola so when plotted on the same graph as $r\kappa$ (solid curve in Fig.~\ref{fig:ambient}) one simply searches for intersection points.  For real metals the Lorenz constant does vary slightly with temperature;  Also, accounting for radiative transfer would effectively increase both $h$ and $L_0$, especially at higher temperatures.  For the sake of simplicity, the parabolic right hand side was assumed for three examples on a 10 $\mu$m long sample, plotted in Fig.~\ref{fig:ambient}.  For an ambient temperature of $T=T_1=65$ K, which lies 8 K below the upper limit $T_U$ for allowable net cooling, very small $W$ will produce a parabola which passes through the point (65 K, -1) and from there heads downwards almost vertically until intersecting the $r\kappa$ curve at a temperature slightly below ambient.  

   For the first example, a bias of 4 mV is applied, still somewhat less than $k_B/e=5.6$ mV.  The situation is much more interesting now with a total of three intersection points (labelled a, b, c), the nearest (a) to the starting point occurring at 59 K, i.e. 6 K of net cooling.  Physically, one would expect this thermodynamic system to reside at this nearest intersection.  As the bias is increased further a tangent curve scenario would produce two intersections and immediately after that just one, so the system would suddenly jump to that remaining intersection point.  The second example describes the situation at a yet higher bias of 20 mV, with the single intersection labelled, d, at a gate temperature of 18.8 K.  Thermodynamic cooling is very strong at this bias level with the system just 0.4 K above the theoretical limit of $T_L=18.4$ K.  One gets a clear sense of how yet larger $W$ would pin the gate temperature very close to $T_L$.  Equation~(\ref{DeltaTind}) would seem to predict a temperature drop greater than $T_1$ with large enough bias, but the more careful analysis leading to Eq.~(\ref{graph}) shows that the gate temperature gets clamped down to $T_L$ at very large $W$, where $P_{cool}$ is only slightly larger than $P_J$.  The last example is the most striking;  Imagine after taking the system to d, the electrical current in the circuit is maintained with 20 mV bias, while the ambient temperature is raised to $T_2$.  Careful inspection of the equations leading to Eq.~(\ref{graph}) that if $T_{GT}$ stays unchanged, then $P_{cool}$ and $P_J$ also stay the same, with only the thermal transfer to the surroundings changing.  This is true even if $T_2$ exceeds $T_U$.  Of course, if $T_2$ is too large with the curvature of the parabola staying constant, eventually the intersection point will be lost and the entire system will relax to ambient temperature.  However, until this happens, TI cooling continues to outpace Joule heating and a type of metastable (stationary) state is maintained.  In the third example $T_2$ is taken to room temperature with $W=20$ mV, and the intersection point, e, occurs at a gate temperature of 20 K. In the laboratory, one could remove all sources of external cooling, let all the visible components warm up to room temperature, and the silver channel would stay at 20 K indefinitely!  
		
	This last example is strong testament to the potency of the type of cooling being discussed here.  The nonequilibrium nature of this thermoelectric system opens up the applied bias as a variable capable of driving, and clamping, the system with tremendous (thermodynamic) force under the right conditions.

\section{Discussion}\label{sec:disc}

From my numerical calculations on highly purified silver and copper, the prospects are very positive for observing TI cooling in the temperature range 15 K to $\approx$200 K, and for sample sizes in the range 400 nm to 1 mm.  It would be desirable to extend this temperature range.  At the upper limit, finding a system with $T_U$ above room temperature would create the convenience of not requiring any other cooling such as liquid refrigerants, so that one could thermoelectrically cool (and clamp down) the channel directly from room temperature.  I would expect similar results for other conventional metals such as gold, chromium, etc. as well as highly doped semiconductors, under degenerate conditions~\cite{Sze}, but the details remain to be worked out.  It may be that systems with very high Debye temperatures will have $T_U$ values above room temperature.  Materials with high Debye temperatures and high mobility with long mean free paths should produce high $T_U$ values without requiring very small sample sizes.  
One system that might meet these requirements is the type of graphene sample that has been fabricated to lengths as short as 5 $\mu$m and operated with very high observed mobilities at 5 K~\cite{Stormer2008}.  Semiconductor systems at very low temperatures are also worth exploring, in particular the two dimensional metals formed as accumulated or inverted layers at interfaces.  These semiconductor heterostructures are known to produce conductive layers with very high mobilities which increase with the lowering of temperature~\cite{Stormer1999}.  The effective purity of these systems is enhanced tremendously by having the dopants located at a distance from the conductive layer.  For those materials that do require very small channels, I note that constructing electronic circuits with general purpose wires having features smaller than 1 mm and even down to near 100 nm in width has become routine and under special circumstances, electrical channels of length 15 nm have been demonstrated~\cite{Gervais2011,Kamp2015}.  

Finding temperature and length ranges for purified silver, over which $r\kappa_{net}$ is less than -1, and with values below -3, is significant (see Fig.~\ref{fig:limits}), but larger magnitudes of $\kappa_{net}$ are desirable. The same level of $r\kappa_{net}$ values is expected for other conventional metals and for degenerate semiconductors the magnitude of $\kappa_{conv}$ is not expected to exceed a value of 3~\cite{Sze}.  	Since larger negative values of $\kappa$ would further enhance the extent of TI cooling, materials exhibiting metal-insulator transitions (MIT) are worth investigation if a phase transition can occur with changing temperature.  Such sharp transitions should produce large values of $\kappa_{conv}$ near the transition temperature.  For example PrNiO$_3$ and NiS$_{1-y}$Se$_y$ both have $\kappa_{conv}$ values near +50 at 95 K and 225 K respectively~\cite{Obradors1993,Tokura1998}.   However these systems have a metallic state at high temperatures which changes to an insulating state at lower temperature where electron-electron interactions lead to blockage of charge flow.  This is the case for most MIT systems but the double-exchange system La$_{1-x}$Sr$_x$MnO$_3$ is an interesting exception~\cite{Tokura1998,Urushibara1995,Kawano1996}.  For x=0.15 the resistivity drops sharply below the transition temperature of 230 K.  Over a modest range in temperature $\kappa_{conv}$ is negative with values near -20.  Materials with large $\kappa_{conv}$ values could increase $T_U$, decrease $T_L$, or make TI cooling more feasible on longer channels that would likely be easier to fabricate.  The caveat with MIT systems is the narrow range of temperature for these transitions, i.e, $|\kappa_{conv}|$ might be large only near the transition. Good candidates for general purpose net TI cooling should have $\kappa_{net}$ values well below -1 over a wide temperature range.  I note though that even with a narrow $T_U-T_L$ range, the induction clamping effect discussed above should still be exploited to hold such systems near the transition temperature even after other forms of sample cooling have been removed.  Superconductors have a transition temperature where the conductivity changes a great deal over small temperature changes, and $\kappa$ will take on extremely large negative values.  Its is difficult to say if the clamping procedure will work for superconductors though since TI cooling requires current flow and the critical current density is easily exceeded in practice.  

On the issue of possible applications of TI cooling, the first conclusion is that such applications are much more likely to be found in small systems.  Fortunately small electronic systems are used widely in modern technology, most notably in microelectronic integrated circuits and optoelectronic devices.  In these systems circuit elements on the 100 nm scale and smaller are routinely implemented.  Moreover, cooling technology that matches these small scales is highly sought after.  Such on-chip Peltier cooling technology exists~\cite{Alley2009} and may be beneficial for devices operating near room temperature.  However, for lower temperature applications this type of cooling is known to lose effectiveness rapidly as the temperature is lowered~\cite{AM}.  It seems then that TI cooling might create a breakthrough in attempts to achieve low temperature thermoelectric cooling of small devices, to temperatures near 20 K or even lower.  High mobility materials are desired for producing fast electronic devices~\cite{Feng2005,Avouris2010} and since mobilities often increase at lower temperatures, TI may hold the key to producing very fast integrated circuitry with built-in cooling.   Elements of any future quantum computers, such as interacting quantum dots would likely require cooling to very low temperatures to maintain coherence~\cite{DiVincenzo1999,Affleck2011}; TI may be used to achieve cooling to temperatures well below 1 K without the need of expensive cryostats.  
The thermodynamically induced temperature clamping effect can also be exploited; Circuits might be cooled by other means initially but after the temperature is clamped TI cooling alone will suffice and sub-circuits running at low temperatures can be incorporated into larger room-temperature circuitry.

A further word is in order on the temperature clamping by TI:  in the example discussed in Fig.~\ref{fig:ambient} the purified silver channel is initially brought down to 65 K by some external and traditional method of cooling.  Then the bias of 20 mV is applied and the temperature is clamped down to 18 K.  Afterwards, the  external cooling is removed and ambient temperature is returned to room temperature.  The channel remains clamped at about 20 K.  If the bias is then turned to zero, the metastable stationary state collapses and one finds oneself with a channel at room temperature, and re-applying will not re-establish the clamp.  One is forced to externally re-cool the channel.  I point out that this state of affairs has much in common with a scheme I have specified for the potential construction of an atomic tether using a scanning tunneling microscope, also exploiting the TI principle.  In that construction, electrical bias is applied to the tunnel junction and the tether is built up one atom at a time.  As the tether grows more applied bias may be required to help stabilize the structure.  If the bias is set to zero, the tether collapses.  Though neither scheme has been realized yet, it is interesting to point out the common features.  In both cases, a thermodynamic state is bolstered by the creation of nonequilibrium conditions (applied bias), and neither would ever arise simply by the application of bias;  The correct sequence of precise steps is required for formation.  Though in both cases this bolstering might at first seem counterintuitive,  at a very basic level these states are stabilized by my postulated statistical weighting factor in Eq.~(\ref{Pi}).  The function that describes a nonequilibrium system's total rate of entropy production would seem to play a vital role.    The realization of one scheme (tether or clamp) would compel one to attempt the other, and search for more examples, as the TI principle seems to have great generality.

\section{Conclusions}\label{sec:Conc}

In conclusion, a method has been developed for achieving a new type of thermoelectric cooling that makes use of the principle of thermodynamic induction. The extent of this thermodynamically-induced cooling is determined by how strongly the electrical conductance depends on temperature.  Such dependence has been investigated for both conventional and ballistic electrical transport.  For ballistic transport a no-net cooling theorem has been established which accounts for induction cooling and Joule heating.  Good candidate nano-systems for observing TI cooling are very short chains, resonant tunneling structures, ribbons, and thick nanosheets.  For conventional transport, investigation of bulk purified silver and copper channels shows that net cooling is possible over wide temperature ranges.  These tests would be conducted for smaller samples, and under higher bias, than in previous transport studies.  Under the right conditions, induction cooling is very potent and can be used to clamp the temperature of a metallic channel in a metastable type of stationary state.  

 A successful search for this inductive cooling would demonstrate the general validity of the induction principle.   In a broader context, a simple statistical weighting factor for nonequilibrium stationary-type inducer states has been developed here, and would seem to have application to many irreversible systems. 
Achieving TI cooling would go a long way towards establishing TI as an essential and general scientific principle. 


\begin{acknowledgements}
I thank Cathy J. Meyer for her assistance in editing the manuscript.
\end{acknowledgements}




\end{document}